\documentclass[conference]{IEEEtran}
\IEEEoverridecommandlockouts
\usepackage{cite}
\usepackage{amsmath,amssymb,amsfonts}
\usepackage{algorithmic}
\usepackage{graphicx}
\usepackage{textcomp}
\usepackage{xcolor}
\usepackage{listings}
\usepackage{hyperref}

\usepackage{subfig}
\usepackage{combelow}

\def\BibTeX{{\rm B\kern-.05em{\sc i\kern-.025em b}\kern-.08em
    T\kern-.1667em\lower.7ex\hbox{E}\kern-.125emX}}

\begin{document}

\title{Large-Scale Graphs Community Detection using Spark GraphFrames}

\author{
    \IEEEauthorblockN{
        Elena-Simona APOSTOL$^{1,2,*}$\thanks{$^*$ These authors contributed equally to this work.},
        Adrian-Cosmin COJOCARU$^{1}$
        and
        Ciprian-Octavian TRUIC\u{A}$^{1,*}$
    }
    \IEEEauthorblockA{
        $^1$ \textit{National University of Science and Technology Politehnica Bucharest, 313 Independen\cb{t}ei, 060042, Bucharest, Romania}\\
        $^2$ \textit{Academy of Romanian Scientists, 3 Ilfov, Bucharest, Romania} \\
        \texttt{
        elena.apostol@upb.ro,
        adrian.cojocaru1603@stud.acs.upb.ro,
        ciprian.truica@upb.ro
        }
    }
}

\maketitle

\begin{abstract}
With the emergence of social networks, online platforms dedicated to different use cases, and sensor networks, the emergence of large-scale graph community detection has become a steady field of research with real-world applications.
Community detection algorithms have numerous practical applications, particularly due to their scalability with data size.
Nonetheless, a notable drawback of community detection algorithms is their computational intensity~\cite{Apostol2014}, resulting in decreasing performance as data size increases.
For this purpose, new frameworks that employ distributed systems such as Apache Hadoop and Apache Spark which can seamlessly handle large-scale graphs must be developed.
In this paper, we propose a novel framework for community detection algorithms, i.e., K-Cliques, Louvain, and Fast Greedy, developed using Apache Spark GraphFrames.
We test their performance and scalability on two real-world datasets.
The experimental results prove the feasibility of developing graph mining algorithms using Apache Spark GraphFrames.
\end{abstract}


\section{Introduction}

Large-scale graphs, i.e., directed or undirected sparse graphs that contain vast amounts of nodes and edges, can be used to model different data collected from online platforms.
Analyzing large-scale graphs can be useful in understanding online platforms to determine user behaviors or research collaborations.

Social media platforms enable users to create and share content, connect with other users, and participate in different events.
Due to the different connections between users, social media data can be modeled as large-scale graphs.
These graph structures model user profiles as the nodes and the relationship, e.g., friendship, likes, shares, follows, etc., between them as the edges.
Thus, different types of graphs can be created using the relationship between nodes, e.g., friendship networks, follower networks, influence propagation, community detection, network immunization, etc.

Research collaborations can also be modeled as large social graphs.
Analyzing and detecting communities within these networks can improve our understanding of how research is conducted and how to determine and propose new collaborations~\cite{Radulescu2020,Radu2020}.
These collaboration networks will model authors as nodes and the co-authors' relationship as an edge between them.

Graph mining can be utilized to analyze large-scale social graphs~\cite{Truica2018,Petrescu2021,Truica2023} in order to extract knowledge and gain a better understanding of user behavior or researcher collaborations.
Community detection is one research direction that enables the identification of significant user interactions on social platforms or researchers' scientific output~\cite{Fortunato2010}, enabling the analysis of how well-connected groups of people interact within a community rather than with the entire social network.
The analysis of communities can help understand how relationships are made and what kind of content and concepts are propagated between the users~\cite{Coban2023}.

The main issue with analyzing these large-scale graphs is their size.
The vast amount of nodes and edges can prove challenging when processing, analyzing, and visualizing these structures when using traditional methods.
To address this issue, we propose a GraphFrame~\cite{Ankur2016} implementation of three community detection algorithms: Louvain, Fast Greedy, and K-Cliques.
We develop the framework on top of the Apache Hadoop ecosystem and Apache Spark~\cite{Zaharia2012}.
The implementation uses YARN~\cite{Vavilapalli2013} as the resource negotiator.

The research question we are trying to answer with this work is: \\
\textit{RQ1:} Can we improve the analysis of large scale graphs using GraphFrames?

The main contribution of our work is a community detection framework developed using GraphFrames on top of the Apache Hadoop ecosystems and Apache Spark.

The rest of this paper is structured as follows. 
Section~\ref{sec:sota} discusses current research related to community detection. 
Section~\ref{sec:methodology} introduces the algorithms and their implementation using GraphFrames. 
Section~\ref{sec:results} presents and discusses the experimental results.
Lastly, Section~\ref{sec:conclusions} presents the conclusions and future research directions.

\section{Related Work}\label{sec:sota}

With the emergence of social networks, online platforms dedicated to different use cases, and sensor networks, the emergence of large-scale graph analytics has become a steady field of research~\cite{Mothe2017}.
New applications that analyze different data types that can be modeled as graphs have been developed in the current years.
These applications have a significant social impact by addressing challenging societal problems, e.g., 
the spread of Covid-19 and epidemic diffusion analysis~\cite{Wickramasinghe2021,Lima2015},
spread of fake news and harmful speech on social networks~\cite{Coban2023},
disaster management and reporting~\cite{Apostol2023,Wilson2016},
deriving socio-economical indicators~\cite{Pappalardo2015,Steele2017},
urban sensing and planning~\cite{Blondel2015,Calabrese2015},
traffic engineering~\cite{Alexander2015,Jarv2012}, 
predicting energy consumption~\cite{Bogomolov2016}, etc. 

The exploration of meaningful communities in graph mining represents a crucial task with significant real-world applications~\cite{Fortunato2010}.
Often, community detection is tackled through the employment of graph clustering algorithms (e.g., K-Cliques~\cite{Palla2005}, etc.) or modularity-based ones (e.g. Louvain~\cite{Blondel2008}, Fast Greedy~\cite{Clauset2004}, etc.).
These algorithms have numerous practical applications, particularly due to their scalability with data size~\cite{Greene2010,Aynaud2010}.
Furthermore, they have been employed in various studies concerning both static and evolving community detection. Nonetheless, a notable drawback of community detection algorithms is their computational intensity~\cite{Stroie2020}, resulting in decreasing performance as data size increases.
For this purpose, new frameworks that can seamlessly handle large-scale graphs must be developed on top of distributed systems such as Apache Hadoop and Apache Spark.


\section{Methodology}\label{sec:methodology}

In this section, we present the community detection algorithms, followed by their implementation using Apache Spark and GraphFrames.

\subsection{Community Detection Algorithms}

We employ three Community Detection algorithms: Lovain, Fast Greedy, and K-Cliques.

\subsubsection{Louvain}
The Louvain algorithm~\cite{Blondel2008} focuses on maximizing the modularity function defined for any undirected graph as follows (Equation~\eqref{eq:modularity}), where  ${A_{i,j}}$ represents the weight of the edge between $i$ and $j$, ${k_i = \sum_{j} A_{i,j}}$ is the sum of the weights of the edges attached to vertex $i$, ${c_i}$ is the community to which vertex $i$ is assigned, the $\delta$-function $\delta(u, v)$ is 1 if $u = v$ and 0 otherwise and ${m = \frac{1}{2} \sum_{ij} A_{i,j}}$, the sum of all the weights of the edges in the graph.

\begin{equation}\label{eq:modularity}
    Q = \frac{1}{2m}\sum_{i,j}\left(A_{i,j} - \frac{k_ik_j}{2m}\right)\delta(c_i,c_j)
\end{equation}

The modularity function is the sum of the comparisons between the weight value of an edge between two vertices $i$ and $j$ (${A_{i,j}}$) and the probability of having such an edge in a random, undirected graph but with the same degree distribution as the original network ($P = \frac{k_ik_j}{2m}$)~\cite{Dugue2022}.
Therefore, a small degree edge between two vertices (${k_i, k_j}$) always contributes more to the modularity function than a high degree one.

For directed graphs, the modularity considered the direction between two nodes $i$ and $j$.
Thus, when $i$ has a large out-degree and a small in-degree and $j$ has a large in-degree and a small out-degree~\cite{Leicht2008}, then the probability of having an edge from $j$ to $i$ is smaller than the probability of having an edge $i$ to $j$.
The in-degree is calculated as the sum of the weights of the edges that go into the node, while the out-degree is computed as the sum of the weights of the edges that go out of the node. 
In the case of directed graphs with random edges but with the same degree distributions, the probability of having an edge between $j$ and $i$ ($j \rightarrow i$) is $P = \frac{k_i^{in}k_j^{out}}{m}$. 
Equation~\eqref{eq:modularity_directed} presents the modularity for a directed graph.
Because the graph is directed, 2 from the denominator is removed.
\begin{equation}\label{eq:modularity_directed}
    Q_d = \frac{1}{m}\sum_{i,j}\left(A_{i,j} - \frac{k_i^{in}k_j^{out}}{m}\right)\delta(c_i,c_j)
\end{equation}

To maximize the modularity and determine communities, the Louvain algorithm performs two steps: 
\begin{enumerate}
    \item[$S_1$:] Each node $i$ is considered a community. The algorithm iterates through all the nodes randomly and computes the improvement in the modularity when assigning nodes to communities. It then chooses the communities with the highest gain. This process is repeated until the modularity stops increasing, i.e., a local maximum has been found. At the end of this phase, each node is assigned to a particular community of nodes. 
    \item[$S_2$:] A meta-network is built based on the existing network and communities found in the previous step. Thus, each community becomes a node (super node), and the weights of the links between the new communities are given by the sum of the weight of the links between nodes enclosed by the two communities. 
\end{enumerate}

For directed graphs, the modularity gain~\cite{Leicht2008} is calculated by adding vertex $i$ to the community $C$ (Equation~\eqref{eq:modularity_gain}), where ${k_i^C}$ is the degree of node $i$ in the community $C$, ${k_i^{out}}$ and ${k_i^{in}}$ are the out-degree and in-degree of node $i$ respectively, and ${\sum_{tot}^{out}}$ and ${\sum_{tot}^{in}}$ are the number of out-going and in-going edges that are incident to community $C$.

\begin{equation} \label{eq:modularity_gain}
    \Delta{Q_d} = \frac{k_i^C}{m} - \left(\frac{k_i^{out}\cdot\sum_{tot}^{in} + k_i^{in}\cdot\sum_{tot}^{out}}{m^2}\right) 
\end{equation}

\subsubsection{Fast Greedy}

Fast Greedy uses the Clauset-Newman-Moore greedy modularity maximization~\cite{Clauset2004} to obtain communities. 
The main objective is the maximization of the same modularity function using a greedy approach.
The method stores in memory the sparse matrix $\Delta{Q}$ that contains the modularity gain $\Delta{Q_{ij}}$ obtained when merging two adjacent nodes (or communities) $i$ and $j$. 

The Fast Greedy uses three steps as follows:
\begin{enumerate}
    \item[$S_1$:] Calculate the initial values of ${\Delta{Q_{ij}}}$ for any adjacent vertices $i$ and $j$ in the graph based on equation \ref{eq:modularity_gain} and populate a max-heap $H$ with the maximum of each row of $\Delta{Q}$.
    \item[$S_2$:] Select the largest element of $H$ (peak) in  ${\Delta{Q_{ij}}}$ and join the corresponding communities, i.e., $i$ merges with $j$.
    $\Delta{Q}$ values are updated to represent neighbors of $i$ or $j$. 
    Also, the values of the max-heap $H$ are updated as well as the values corresponding to the heaps for $i$ and $j$ neighbors.
    \item[$S_3$:] Repeat Step $S_2$ until only one community remains.
    When two communities $i$ and $j$ are joined the algorithm will update the $j$-th row of matrix $\Delta{Q}$ and remove the $i$-th row completely. 
\end{enumerate}

For any of $j$'s neighbors $q$, the updates for ${\Delta{Q_{jq}}}$ is 
$\Delta{Q'_{jq}} = \Delta{Q_{iq}} + \Delta{Q_{jq}}$.
When $q$ is connected to both $i$ and $j$, when merging $i$ and $j$ in an undirected graph the update of ${\Delta{Q_{jq}}}$ is $\Delta{Q'_{jq}} = \Delta{Q_{iq}} - \frac{k_jk_q}{2m^2}$, while for directed graphs is $\Delta{Q'_{jq}} = \Delta{Q_{iq}} -\frac{k_j^{in}k_q^{out} + k_j^{out}k_q^{in}}{2m^2}$.
When $q$ is connected to $i$ and not $j$, the update for an undirected graph is $\Delta{Q'_{jq}} = \Delta{Q_{jq}} -  \frac{k_ik_q}{2m^2}$, while for a directed graph is $\Delta{Q'_{jq}} = \Delta{Q_{jq}} -  \frac{k_i^{in}k_q^{out} + k_i^{out}k_q^{in}}{2m^2}$.

The modularity $Q$ will reach one single maximum value, since after the largest $\Delta{Q}$ (i.e., peak of heap $H$) becomes negative, all the other $\Delta{Q}$ decrease. 

\subsubsection{K-Cliques}

The K-Cliques communities algorithm assumes that a community consists of several fully connected subgraphs that contain common nodes~\cite{Palla2005} and creates communities as unions of all K-Cliques (complete subgraphs of size $k$) that can be reached from each other through a series of adjacent K-Cliques, where adjacency means sharing $k-1$ nodes.
The algorithm takes advantage of nodes that are members of multiple well-connected subsets of nodes to determine overlapping communities and manages to merge smaller cliques (i.e., number of nodes less than $k$) into larger communities using these overlapping community nodes.

\subsection{Implementation}~\label{ssec:implementation}

Our implementation uses Apache Spark and the GraphFrames API.
The GraphFrames API is based on the Spark DataFrame API.
A DataFrame is a table structure that contains data grouped into columns and rows.
A GraphFrame consists of two DataFrames, one for edges and one for vertices, and allows working with graphs using 2-dimensional representations, i.e., tables.
Furthermore, a GraphFrame can be used with graph-specific functions and operations as well as with all DataFrame-specific functions and operations.
The code is available on GitHub at \url{https://github.com/DS4AI-UPB/CommunityDetection-DataframesSpark}.

\section{Experimental Results}~\label{sec:results}

In this section, we present the datasets used for the experiments, the experimental setup, and the time performance and scalability tests. 

\subsection{Datasets details}~\label{sec:dataset}

For our experiments, we use a Twitter (currently X online platform) dataset and Research Collaboration datasets.

\begin{figure}[!htb]
    \centering
    \subfloat[Tweet\label{fig:tweet}]{{\includegraphics[width=0.75\columnwidth]{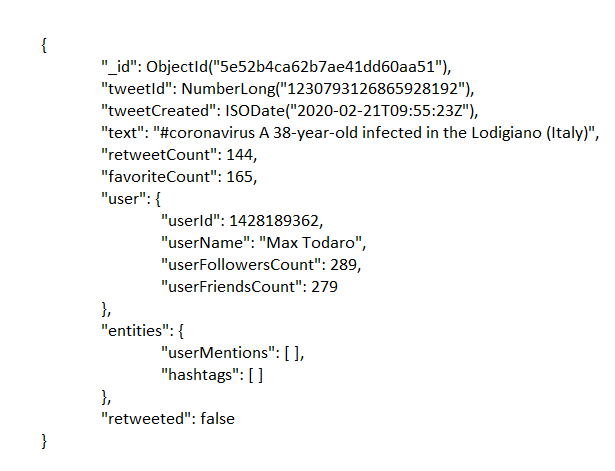}}}
    \hfill    
    \subfloat[Retweet\label{fig:retweet}]{{\includegraphics[width=0.75\columnwidth]{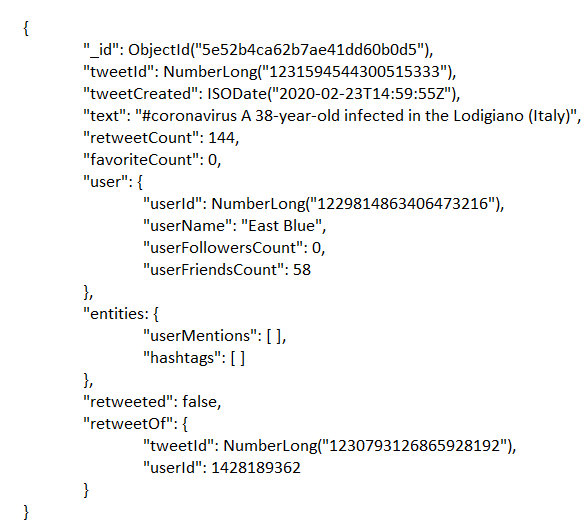}}}
    \caption{Record Examples}
    \label{fig:tweet_retweet}
\end{figure}

The Twitter dataset consists of $50\,230$ tweets and retweets posted during February and March 2020, from the territory of Italy, with the onset of the crisis caused by the appearance of Covid-19 virus infection.
These tweets were collected using Twitter API and the information was retrieved in JSON format.
Figure~\ref{fig:tweet} presents a tweet record, while Figure~\ref{fig:retweet} presents a retweet example.

We build a graph $G = (E, V)$ using users, tweets, and retweets.
A user becomes a node if he/she posts a tweet or retweets one.
The vertices in our graph are represented by retweets.
If a user retweets another user's tweet, we create a vertex with weight 1 between these users.
Whenever we encounter another retweet relationship between these two users, we increase that weight (+1).

The Research Collaborations graph consists of $\sim35\,000$ authors (nodes) that have between them $\sim500$ collaborations (edges).

\subsection{Experimental setup}

For our experiments, we used virtual machines provided by  Databricks (Table~\ref{tab:vms}).

\begin{table}[!htbp]
    \centering
    \caption{Databriks virtual machines}\label{tab:vms}
    \begin{tabular}{lrr}
        \hline
        \textbf{VM}        & \textbf{RAM}  & \textbf{Cores} \\ 
        \hline
        m4.large  &  8GB &  2    \\
        m4.xlarge & 16GB & 16    \\
        i3.xlarge & 32GB &  4    \\
        \hline
    \end{tabular}
\end{table}

\begin{figure}[!htb]
\centering
\includegraphics[width=0.8\columnwidth]{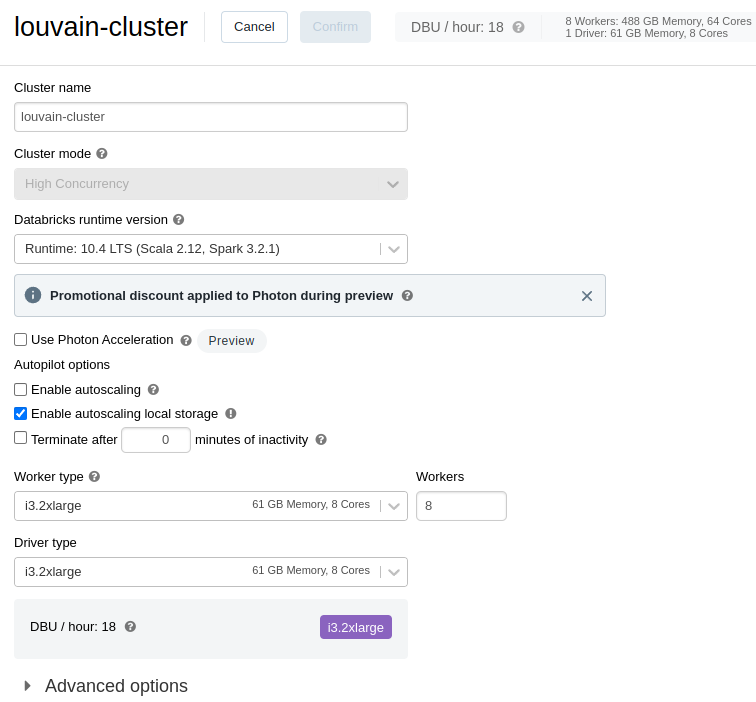}
\caption{Databricks Spark VM configuration example}
\label{fig:louvain-cluster} 
\end{figure}

All the virtual machines have the same software configurations (Figure~\ref{fig:louvain-cluster}): Scala 2.12 and Spark 3.2.1.
We also installed the GraphFrame package that works with the given version of Scala. 
Finally, each algorithm was executed on its own virtual machine.

\subsection{Results}

In this section, we present the obtained results. 
For the time performance of the algorithms, we use the Twitter dataset with different environments.
For the scalability experiments, we use the Research Collaborations dataset.

\subsubsection{Time performance}

For this set of experiments, we use the Twitter dataset.

Table~\ref{tab:louvain_full_ds} presents the performance of Louvain on the full Twitter dataset using different environments.
We can observe that, depending on the configurations the execution time is almost constant.
These results are expected as, with the increase in the number of workers, the job initialization increases as well, as it is the most time-consuming operation on a Spark Cluster~\cite{Truica2020}.

\begin{table}[!htpb]
\caption{Louvain Experiments}\label{tab:louvain_full_ds}
\centering
\begin{tabular}{llrr}
\hline
\textbf{Environment} & \textbf{VM} & \textbf{No. Workers} & \textbf{Time (s)} \\
\hline
Databricks & i3.xlarge & 1 & 1\,280.56s \\
Databricks & i3.xlarge & 2 & 1\,530.57s \\
Databricks & i3.xlarge & 4 & 1\,660.49s \\
Databricks & m4.xlarge & 2 & 1\,440.44s \\
Databricks & m4.xlarge & 4 & 1\,518.93s \\
Databricks & m4.xlarge & 8 & 1\,590.71s \\
\hline
\end{tabular}
\end{table}

We use the same approach to test the time performance of the Fast Greedy algorithm on the full Twitter dataset.
Table~\ref{tab:fastgreedy_full_ds} presents the obtained experimental results.
We observe again that the performance is almost constant due to the same reasons as for Louvain.
The experiment shows that Fast Greedy is faster than Louvain, this is in part due to the way the maximum modularity is computed.

\begin{table}[!htpb]
\caption{Fast Greedy Experiments}\label{tab:fastgreedy_full_ds}
\centering
\begin{tabular}{llrr}
\hline
\textbf{Environment} & \textbf{VM} & \textbf{No. Workers} & \textbf{Time (s)} \\
\hline
Databricks & i3.xlarge & 1 & 352.46s \\
Databricks & i3.xlarge & 2 & 339.88s \\
Databricks & i3.xlarge & 4 & 333.42s \\
Databricks & m4.xlarge & 2 & 330.06s \\
Databricks & m4.xlarge & 4 & 412.43s \\
Databricks & m4.xlarge & 8 & 401.17s \\
\hline
\end{tabular}
\end{table}

\subsubsection{Scalability testing}

For this set of experiments, we use the Research Collaboration dataset.
We split the dataset using three different scale factors.
We start with a small graph with only 352 authors and 60 articles and increase both the number of nodes as well as the number of edges to 34\,986 and 500, respectively.
We also experiment with different workers on the same virtual machine, i.e., m4.large. 

The first set of experiments is for the Louvain algorithm (Table~\ref{tab:louvain_sc}).
We observe that the time performance increases with the graph size.
Moreover, the time performance slightly decreases linearly when the number of workers increases. 
We can conclude that the algorithms' time performance is directly impacted by the workers' number of cores. 
Thus, the time performance decreases when using more workers than cores because each machine has only 2 cores.

\begin{table}[!htpb]
\centering
\caption{Louvain scalability results using the Research Collaborations graph}
\label{tab:louvain_sc}
\begin{tabular}{rrrrr}
\hline
Edges & Vertices & 1 Worker & 2 Worker & 4 Worker \\
\hline
    352 &  60 &  89.66s  & 102.85s & 105.48s \\
 8\,810 & 250 & 384.77s  & 350.70s & 362.62s \\
34\,986 & 500 & 804.09s  & 679.03s & 733.21s \\ 
\hline
\end{tabular}
\end{table}

Table~\ref{tab:greedy_sc} presents the results for the Fast Greedy community detection algorithm.
Although the algorithm converges faster than Louvain, we observe that the scalability remains limited by the amount of resources used. 

\begin{table}[!htpb]
\centering
\caption{Fast Greedy scalability results using the Research Collaborations graph}
\label{tab:greedy_sc}
\begin{tabular}{rrrrrr}
\hline
Edges & Vertices & 1 Worker & 2 Worker & 4 Worker\\
\hline
    352 &  60 &  17.11s &  17.52s &  17.76s  \\
 8\,810 & 250 &  69.00s &  74.93s &  97.86s  \\
34\,986 & 500 & 177.87s & 168.43s & 164.66s  \\ 
\hline
\end{tabular}
\end{table}

Table~\ref{tab:cliques_sc} presents the results for K-Cliques.
From a scalability perspective, this algorithm falls between Fast Greedy and Louvain.
We note that, due to the heuristic used to determine communities, the performance is less impacted by the resources allocated to the virtual machine.

\begin{table}[!htpb]
\centering
\caption{K-Cliques scalability results using the Research collaborations graph}
\label{tab:cliques_sc}
\begin{tabular}{rrrrrr}
\hline
Edges & Vertices & 1 Worker & 2 Worker & 4 Worker\\
\hline
    352 &  60 &    6.55s &    6.29s &   5.67s   \\
 8\,810 & 250 &  192.91s &  162.46s & 144.52s   \\
34\,986 & 500 &  297.52s &  248.84s & 172.72s   \\ 
\hline
\end{tabular}
\end{table}

\section{Conclusions}~\label{sec:conclusions}

In this article, we presented a large-scale graph community detection framework developed on top of Spark GraphFrames to answer \textit{RQ1}.
We implemented three community detection algorithms, i.e., Louvain, Fast Greedy, and K-Cliques.
The experimental results show that the algorithms have good performance and linear scalability as they take advantage of the in-memory distributed processing capabilities of Spark and the 2-dimensional table representation of data.
In future work, we aim to improve our framework for large-scale graph analysis with Infomod~\cite{Rosvall2008} and Infomap~\cite{Rosvall2010}.

\section*{Acknowledgment}

This work is supported in part by
(1) The National University of Science and Technology Politehnica Bucharest through the ``PubArt'' project;
(2) The German Academic Exchange Service (DAAD) through the project ``iTracing: Automatic Misinformation Fact-Checking'' (DAAD grant no. 91809005); and
(3) The Academy of Romanian Scientists through the funding of ``SCAN-NEWS: Smart system for deteCting And mitigatiNg misinformation and fake news in social media''.

\bibliography{main}
\bibliographystyle{IEEEtranS}

\end{document}